# Equations for filling factor estimation in opal matrix

S. M. Abrarov[a], T. W. Kim[b], T. W. Kang[a]

[a]*Quantum-functional Semiconductor Research Center, Dongguk University, Seoul 100-715, South Korea*

[b]*Advanced Semiconductor Research Center, Division of Electrical & Computer Engineering, Hanyang University, Seoul 133-791, South Korea*

**Abstract**

We consider two equations for the filling factor estimation of infiltrated zinc oxide (ZnO) in silica (SiO$_2$) opal and gallium nitride in ZnO opal. The first equation is based on the effective medium approximation, while the second one – on Maxwell-Garnett approximation. The comparison between two filling factors shows that both equations can be equally used for the estimation of the quantity of infiltrated nanocrystals inside opal matrix [⌘].

## I. Introduction

Photonic crystals (PhCs) with forbidden band-gaps, proposed by Eli Yablonovitch[1] and Sajeev John[2], open new opportunities for their applications in modern optics. PhCs are one-, two-, and three-dimensional dielectric lattices with periodicity on the order of the optical wavelengths. The implementations of PhCs are mostly aimed to improve the useful properties of various materials as well as opto-electronic devises such as light emitting diodes[3], laser diodes[4], optical fibers[5]. Nowadays research on PhCs becomes an increasingly important in the fundamental and applied sciences.

One of the kinds of PhCs is an opal matrix consisting of spherical sub-micron balls packed into face centered cubic (FCC) structure by means of self-sedimentation in a fluid suspension[6,7]. Silicon dioxide (SiO$_2$) or silica is frequently used as a host material in artificial opals. Silica balls are synthesized by Stöber-Fink-Bohn process through the hydrolysis of tetraethylorthosilicate in the ethanol solution mixed with ammonium hydroxide and water[8].

The applications of opal PhCs have number of significant advantages over others. For instance, the opal matrix can be grown over a large practically unlimited plane area. Their

---

[⌘]**Addendum to the article** *http://arxiv.org/abs/physics/0508152*

fabrication is very technological without requirement for expensive equipment. It has been recently shown that by means of electro-deposition the high quality two- and three-dimensional porous films, patterned in inverted opal, can be successfully realized [9]. Thus, the nanocrystals grown in artificial opal can be regarded as inexpensive and efficient alternative for electro- and photolithography.

The fabrication of high quality artificial opal by natural self-sedimentation in monodispersed fluid suspension may continue for a long period, up to ten months [6]. However this drawback is resolved in electrophoretically assisted sedimentation involving an external electric filed. Such an original technology enables one to accelerate a sedimentation velocity up to 0.2-0.7 mm per hour for the balls ranging in diameter between 300 - 550 nm [10].

By means of the various chemical depositions, the voids of the opal matrix can be filled with semiconductors (GaAs, CdS, HgSe, Si, InN/GaN, CdTl, InP, ZnO, ZnS), superconductors (In, Pb) ferromagnetic materials (Fe and alloys) [11]. Different infiltration methods including chemical vapor deposition [12,13], chemical bath deposition [12], hydrolysis [12], salt-precipitation [13], sol-gel [13,14], electro-deposition [9,13], spray pyrolysis [13,15,16], etc. can be applied for formation of nanoparticles in interglobular spaces of opal matrix.

Figure 1 shows an experimental set for observation of Bragg reflection from the surface of opal matrix with perfectly assembled FCC structure. The reflection peak of the light is governed by Bragg's law

$$\lambda = 2d_{h,k,l}\left(n_{eff}^2 - \sin^2\theta\right)^{1/2} \quad (1)$$

where $\lambda$ is the wavelength,

$$d_{h,k,l} = \frac{a}{\sqrt{h^2 + k^2 + l^2}},$$

$a$ is the distance between planes, $h$, $k$, $l$ are Miller indices, and $n_{eff}$ is the effective refractive index. For opal with perfectly ordered balls, the experimental data excellently fit Bragg's law. Peak in reflectance (or dip in transmittance) shifts to the blue spectrum with increasing angle according to (1). However, an opal with imperfectly ordered silica balls behaves differently.

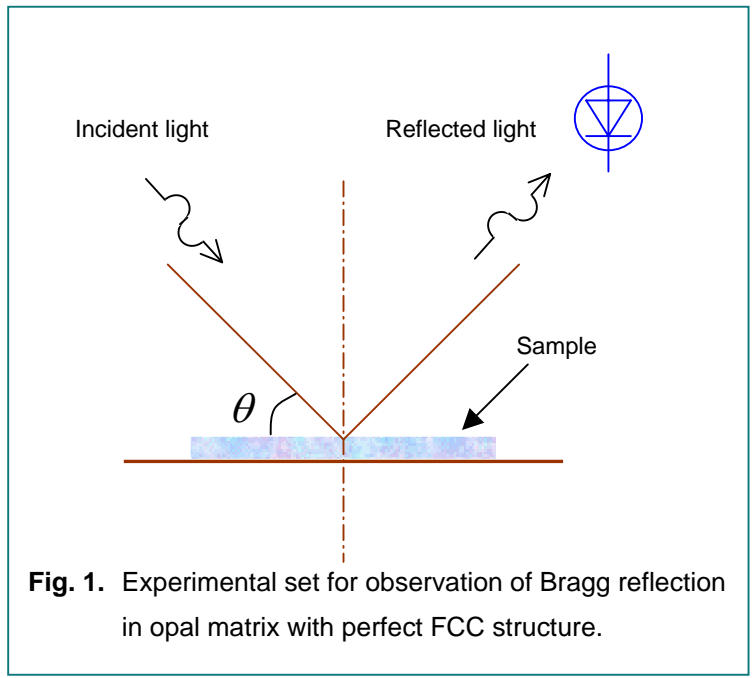

**Fig. 1.** Experimental set for observation of Bragg reflection in opal matrix with perfect FCC structure.

Consider Figure 2 showing the SEM image of the opal, which FCC structure has dislocations, formed during the self-sedimentation process in a fluid suspension. The sample contains micro-size domains with facing up (111) and (100) planes. These domains are chaotically distributed within the sample and differently tilted with respect to its surface. As a result, the blue-shift in reflectance (or in transmittance) becomes insignificant and irregular with increasing angle [17, 18]. It signifies that in highly imperfect or in amorphous opal the color remains practically stable at any $\theta$ (Figure 1). Despite of the fact that such sample does not exhibit the blue-shift with increasing angle, the influence of photonic band-gap (PBG) in opal with disordered FCC structure is possible to observe conclusively either in evolution of photoluminescence arising due to gradual increase of the filling factor or in temperature-dependent photoluminescence [16].

It is worth remarking that imperfect opal structures exhibit spectrum with greater FWHM in reflection (or transmission) [19]. While perfectly ordered opal matrix may find its applications in various light emitting devices [3, 4], imperfectly assembled and/or amorphous opal with embedded luminescent nanoparticles might be useful for applications in full-color displays [16].

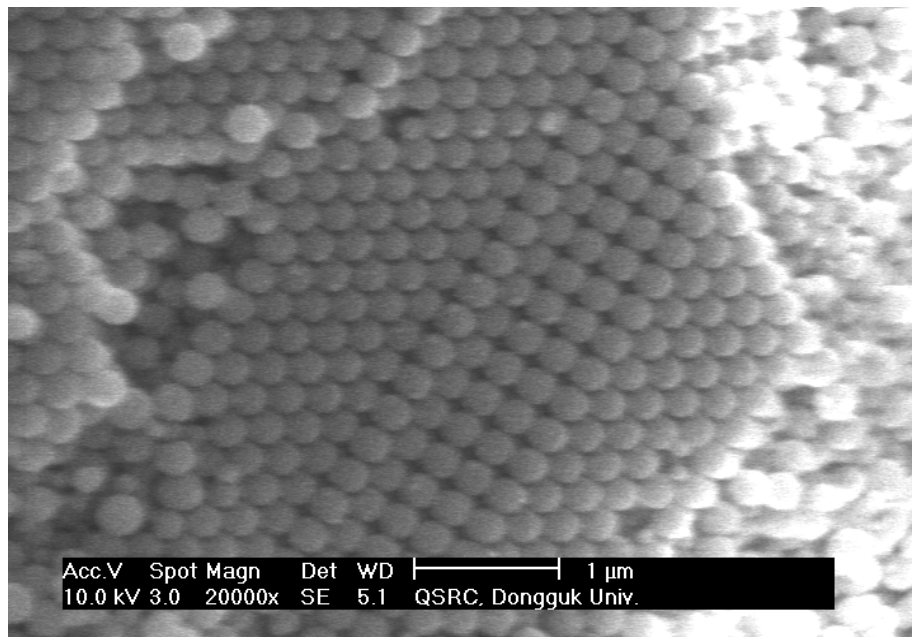

**Fig. 2.** SEM image of the opal comprising silica balls. Average diameter of spheres is around 260 nm **(**orange opal).

Nanoparticles infiltrated in interglobular spaces can considerably alter the optical properties of opal matrix. Therefore the estimation of the quantity of infiltrated nanocrystals plays significant role in practice. Particularly, the amount of infiltrated material has to be properly controlled during its deposition in the voids.

This paper reports two filling factor equations based on effective medium and Maxwell-Garnett approximations. The comparison between them shows that both equations can be equally used for quantity estimation of infiltrated nanoparticles in interglobular spaces between FCC packed spheres.

## II. Equations for filling factor estimation

### 2.1 Refractive index based on effective medium approximation

The quantitative analysis of the optical characteristics of opal matrix can be significantly simplified introducing the effective refractive index according to effective medium approximation[20]. Effective refractive index can be defined as a weighted sum of indices of refraction $n_1$, $n_2$, $n_3$, for spherical balls, infiltrated nanoparticles and air, respectively. For the bare and infiltrated opals, the effective refractive indices accordingly are

$$n_{eff\,1} = n_1 0.74 + n_3 0.26 \qquad (2a)$$

and

$$n_{eff\,2} = n_1 0.74 + n_2 f + n_3 (0.26 - f), \qquad (2b)$$

where $f$ is the filling factor for infiltrated nanocrystals. The values 0.74 and 0.26 are the filling factors for the host material (spherical balls) and air, respectively.

It is convenient to assume a low angle of incidence (Figure 1). Substitution of definitions (2a, b) into (1) yields wavelengths $\lambda_1$, $\lambda_2$ for the bare and infiltrated opals. The refractive indices for the spherical balls and infiltrated nanoparticles are both, in general, wavelength dependent.

The red-shift in reflectance or transmittance spectra arising due to infiltrated nanocrystals inside opal matrix can be found as

$$\Delta\lambda = \lambda_2 - \lambda_1 = 2d_{h,k,l}\left(n_{eff\,2} - n_{eff\,1}\right). \qquad (3)$$

Substituting definitions (2a, b) into (3) yields the relation for the filling factor

$$f = \frac{\dfrac{\Delta\lambda}{2d_{h,k,l}} - (n_1(\lambda_2) - n_1(\lambda_1))0.74}{n_2(\lambda_2) - n_3}. \qquad (4)$$

Alternatively, the filling factor can be derived through ratio between wavelengths $\lambda_1$, $\lambda_2$ for the bare and infiltrated opals

$$\frac{\lambda_2}{\lambda_1} = \frac{2d_{h,k,l}\, n_{eff\,2}}{2d_{h,k,l}\, n_{eff\,1}}. \qquad (5)$$

Substitution of effective refractive indices (2a, b) into (5) gives

$$f = \frac{\dfrac{\lambda_2}{\lambda_1}(n_1(\lambda_1)0.74 + n_3 0.26) - (n_1(\lambda_2)0.74 + n_3 0.26)}{n_2(\lambda_2) - n_3}. \qquad (6)$$

Clearly that (4) and (6) are equivalent. Substitution of expression $\lambda_1 = 2d_{h,k,l}\, n_{eff\,1}$ into (6) leads to (4).

## 2.2 Effective refractive index based on Maxwell-Garnett approximation

Another definition for effective refractive indices, also widely used in practice, is based on Maxwell-Garnett approximation [21]. Effective refractive indices for the bare and infiltrated opals can be expressed as a weighted sum of the squared refractive indices

$$n_{eff\,1}^2 = n_1^2 0.74 + n_3^2 0.26, \qquad (7a)$$

$$n_{eff\,2}^2 = n_1^2 0.74 + n_2^2 f + n_3^2 (0.26 - f). \qquad (7b)$$

Assume again a low angle of incidence. Substitution of definitions (7a, b) into (1) provides two squared wavelengths $\lambda_1^2$ and $\lambda_2^2$ corresponding to the bare and infiltrated opals, respectively. The difference between them is

$$\lambda_2^2 - \lambda_1^2 = (2d_{h,k,l})^2 (n_{eff\,2}^2 - n_{eff\,1}^2). \qquad (8)$$

From (7a, b) and (8) the filling factor can be found as

$$f = \frac{\dfrac{\lambda_2^2 - \lambda_1^2}{(2d_{h,k,l})^2} - (n_1(\lambda_2)^2 - n_1(\lambda_1)^2)0.74}{n_2(\lambda_2)^2 - n_3^2}. \qquad (9)$$

Alternatively, the filling factor can be derived through following fraction

$$\frac{\lambda_2^2}{\lambda_1^2} = \frac{4 d_{h,k,l}^2 \, n_{eff\,2}^2}{4 d_{h,k,l}^2 \, n_{eff\,1}^2}. \qquad (10)$$

Substituting (7a, b) into (10) leads to the relation for the filling factor

$$f = \frac{\frac{\lambda_2^2}{\lambda_1^2}\left(n_1(\lambda_1)^2 0.74 + n_3^2 0.26\right) - \left(n_1(\lambda_2)^2 0.74 + n_3^2 0.26\right)}{n_2(\lambda_2)^2 - n_3^2}. \qquad (11)$$

Obviously (9) and (11) are equivalent. Substitution of expression $\lambda_1^2 = 4 d_{h,k,l}^2 \, n_{eff\,1}^2$ into (11) leads to (9).

Equations (4), (6), (9), and (11) contain wavelength dependent terms $n_1(\lambda)$ and $n_2(\lambda)$. In order to represent them in analytic form, it is convenient to use Sellmeier dispersion formula providing excellent match for ZnO and SiO$_2$ [22, 23, 24]

$$n(\lambda)^2 = A + \frac{B\lambda^2}{\lambda^2 - C^2} + \frac{D\lambda^2}{\lambda^2 - E^2} \qquad (12)$$

where $A$, $B$, $C$, $D$ and $E$ are adjustable characteristics parameters.

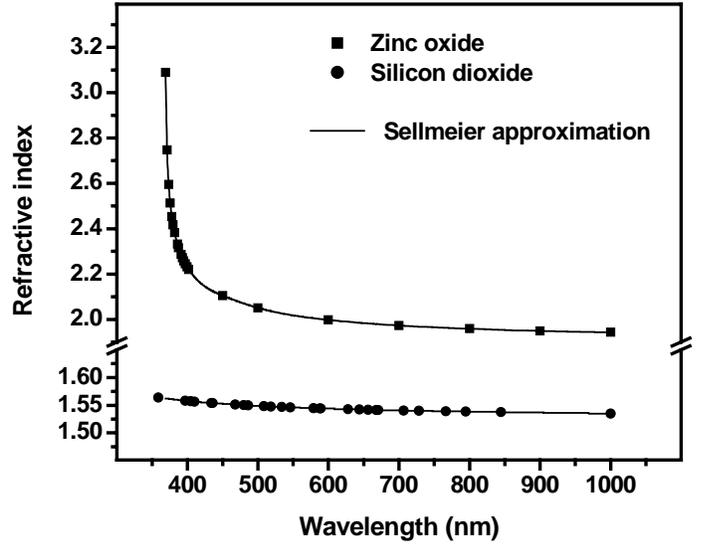

**Fig. 3.** Refractive indices for silicon dioxide and zinc oxide.

## III. Filling factors for opal matrix comprising silica balls

### 3.1 Refractive indices of silicon dioxide and zinc oxide

Figure 3 shows the refractive indices for silicon dioxide and zinc oxide vs. wavelength. Analytical form of $n_1(\lambda)$ and $n_2(\lambda)$, obtained via Sellmeier dispersion approximation (12), quite accurately fit data available in literature [25, 26]. The refractive index of ZnO may be considered a flat only at the wavelengths above 450 nm where in the most of the visible range it is equal to around 2. Below this point, the refractive index of ZnO has strong wavelength dependence and its curve rapidly rises due to resonance occurring between valence and conduction bands.

Contrarily, the curve for the silicon dioxide is nearly flat and consequently its refractive index can be considered a constant over the wide optical range covering near infrared (IR) to near ultraviolet (UV) spectra. Taking this into account and considering the fact that refractive index of air is very close to unity, (4) and (6) can be simplified and approximated as

$$f \approx \begin{cases} \dfrac{\Delta\lambda}{2d_{h,k,l}(n_2(\lambda_2)-1)} \\ \\ \dfrac{\Delta\lambda}{\lambda_1}\dfrac{n_1 0.74+0.26}{n_2(\lambda_2)-1} \end{cases} \qquad (13)$$

Similarly (9) and (11) can also be simplified and represented in form

$$f \approx \begin{cases} \dfrac{\lambda_2^2-\lambda_1^2}{4d_{h,k,l}^2(n_2(\lambda_2)^2-1)} \\ \\ \dfrac{\lambda_2^2-\lambda_1^2}{\lambda_1^2}\dfrac{n_1^2 0.74+0.26}{n_2(\lambda_2)^2-1} \end{cases} \qquad (14)$$

Suppose that the sample shown in the Figure 1 has the (111) plane facing up. In this case

$$d_{1,1,1} = \dfrac{\sqrt{2}D}{\sqrt{3}} \approx 0.816D,$$

where $D$ is the average spherical diameter. Substitution of the value $d_{1,1,1}$ into upper form of (13) results

$$f \approx \dfrac{\Delta\lambda}{2D \times 0.816(n_2(\lambda_2)-1)}. \qquad (15)$$

Equation (15) has been used for ZnO filling factor estimation in our previous work[27].

### *3.2 Discrepancies between filling factors*

The relative error defined as

$$err = \dfrac{|f_1-f_2|}{f_1} \times 100\% \qquad (16)$$

is used in the present work to evaluate discrepancies between filling factors. Figure 4 shows the filling factors and the their relative errors vs. red-shift. The filling factors $f_1$ and $f_2$, calculated according to (13) and (14), are shown by solid and dashed curves, respectively.

Consider silica opals, which PBGs include near UV, visible (purple-blue, bluish-green, green, yellowish-orange, orange-red, crimson) and near IR spectra. Near origin the variations of the filling factors $f_1$, $f_2$ are comparatively high and their relative error strongly depends on diameter of silica balls. For UV opal with $D = 170$ nm the relative error is almost 10%, while for IR opal with $D = 310$ nm it is less than 4%. For all other opals with small infiltrations the relative errors are less than 8%.

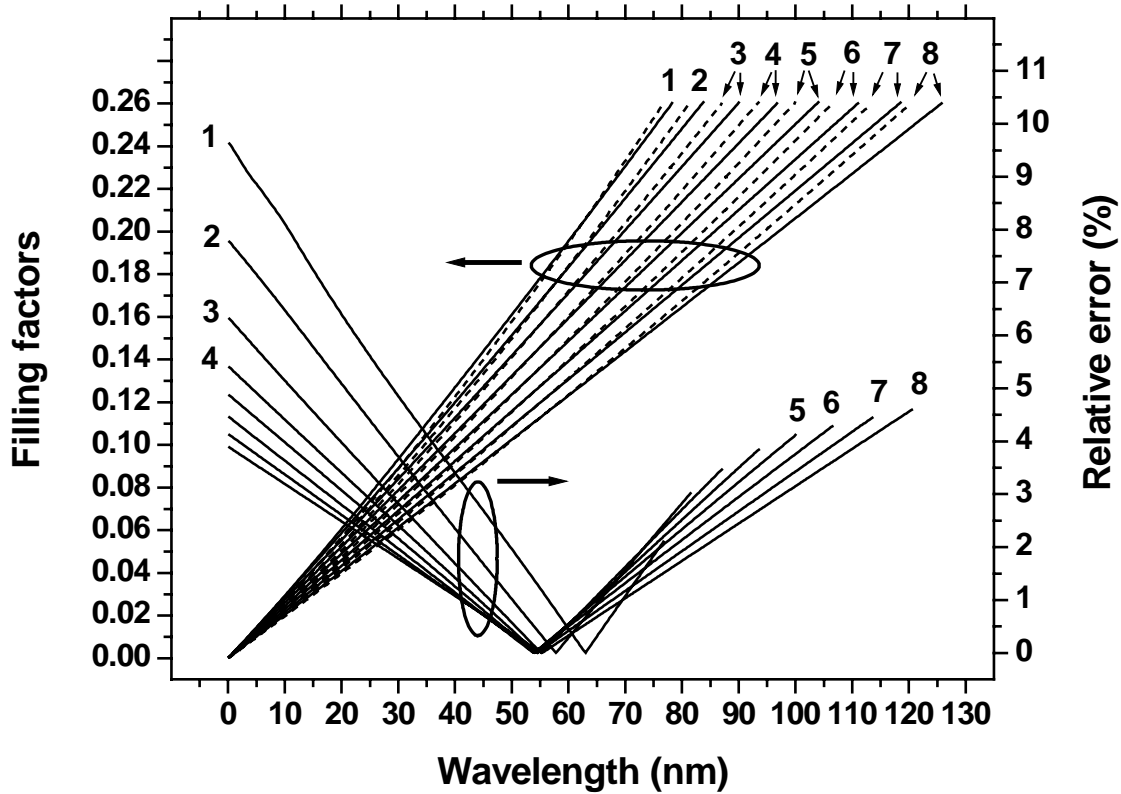

**Fig. 4.** Filling factors and relative error vs. red-shift for silica opals:

**1** – 170 nm (near UV)    **5** – 250 nm (yellowish-orange)
**2** – 190 nm (purple-blue)  **6** – 270 nm (orange-red)
**3** – 210 nm (bluish-green) **7** – 290 nm (crimson)
**4** – 230 nm (green)      **8** – 310 nm (near IR).

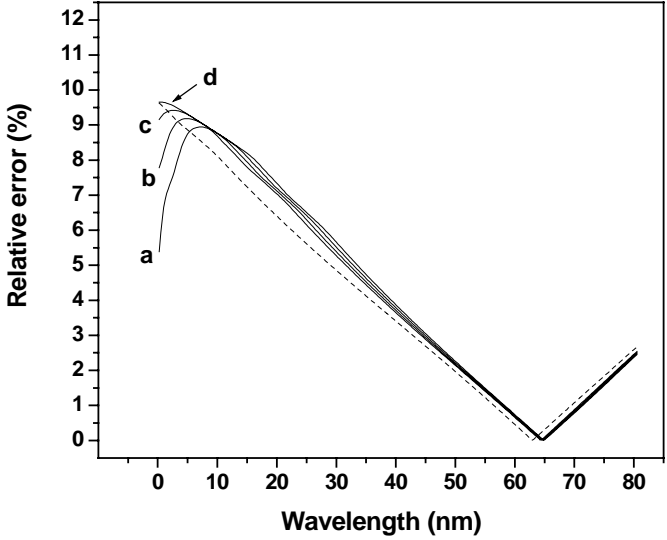

**Fig. 5.** Relative error vs. wavelength for UV opals.

Filling factors $f_1$, $f_2$ intercept each other in the range between 55 to 65 nm indicating that the least discrepancies between filling factors occured when $0.1 \leq f \leq 0.2$.

Figure 5 shows relative error vs. wavelength dependencies for UV opals with average ball diameters: (a) 162, (b) 163, (c) 164, and (d) 165 nm. The dashed curve corresponding to $D = 170$ nm is also shown for comparison. From Figure 5 one can see that the relative error does not further increase with decreasing average diameter of balls.

At the high infiltrations the relative errors are negligible for UV and purple-blue opals. For all other opals they do not exceed 5%. Discrepancies between filling factors show that each of two simplified equations (13) and (14) can be used for the quantity estimation of infiltrated material inside silica opal. However, it should be taken into account that for the small amount of infiltration in UV opal the discrepancy may be relatively high, nearly 10%.

## IV. Filling factors for opal matrix comprising ZnO balls

In fact, both equations (2a) and (7a) contain term $n_1$, which is itself, generally, may depend on value $\lambda_1$. Therefore when the refractive index of balls is a function of the wavelength, either of two equations (2a), (7a) contains two unknowns, namely $\lambda_1$ and $n_1(\lambda_1)$. This problem cannot be resolved analytically due to complicated form of Sellmeier dispersion formula (12). Iterating loop [28] is a useful and efficient programming method to solve numerically such a task. The basic objective in computation is to determine $\lambda_1$ and $n_1(\lambda_1)$, given by (2a) and (7a) via (12). Having known the exact values $\lambda_1$ and $n_1(\lambda_1)$, the filling factors $f_1(\Delta\lambda)$, $f_2(\Delta\lambda)$ can be readily found through corresponding equations (4) and (9), respectively.

The novel approach in fabrication of artificial opal comprising ZnO balls has been reported recently [19]. Refractive index of ZnO in the near UV spectrum is very high, exceeding 9 at the band edge [25, 26]. Therefore, having such a high value of the refractive index, ZnO might be regarded a possible candidate in fabrication of opal matrix with complete PBG.

Gallium nitrate (GaN) can be synthesized inside the voids of opal matrix by means of chemical deposition, which details described elsewhere [29]. Suppose that GaN is infiltrated in ZnO opal. In such a combination $n_1$ and $n_2$ are refractive indices for ZnO balls and GaN, respectively. Unlike silica, ZnO is strongly wavelength dependent in the near UV region. Due to this reason, simplified equations (13) and (14) cannot be applied for opal comprising ZnO balls when $\lambda_1 < 450$ nm.

The algorithm for computation of $\lambda_1$ and $n_1(\lambda_1)$ is straightforward. Consider $\lambda_1$ and $n_1(\lambda_1)$, related to effective medium approximation. Choose an arbitrary trial value of $\lambda_1$, say 500 nm, and include it into Sellmeier dispersion formula (12). Find the corresponding refractive index of balls $n_1(\lambda_1)$ and substitute it into (2a). Calculate $\lambda_1$ and compare it with previous value. If the difference between them is large, include the recent value $\lambda_1$ into Sellmeier dispersion formula and repeat all calculations again. Continue the same procedures if the difference between the recent and previous values of $\lambda_1$ is not greater than some small epsilon, say $10^{-3}$ nm.

The computation of $\lambda_1$ and $n_1(\lambda_1)$, related to Maxwell-Garnett approximation, is absolutely similar with the only difference that it employs (7a) instead of (2a). Typically the iteration consisting of just 20-40 calculation cycles (steps) is sufficient to get a required precision.

Table 1 shows the intermediate results for the opal with zinc oxide balls, which average diameter supposed to be equal to 130 nm. The last row shows the exact values of $\lambda_1$ and $n_1(\lambda_1)$. The right part of Table 1 converges faster to the desired values due to squared form of (7a).

Figure 6 shows filling factors and their relative error for ZnO opal infiltrated with GaN. The filling factors do not intercept. The curve for $f_2$ grows faster than that for $f_1$, consequently the relative error monotonically increase. At the origin the relative error is small, less than 6%. However, at complete infiltration the discrepancy between filling factors becomes relatively high, reaching almost 16%.

| | *Effective medium approximation* | | | | *Maxwell-Garnett approximation* | | |
|---|---|---|---|---|---|---|---|
| **Step** | Trial $\lambda_1$, nm | Calculated $\lambda_1$, nm | $n_1(\lambda_1)$ | **Step** | Trial $\lambda_1$, nm | Calculated $\lambda_1$, nm | $n_1(\lambda_1)$ |
| 1 | 500.000 | 377.395 | 2.051 | 1 | 500.000 | 389.878 | 2.051 |
| 2 | 377.395 | 442.312 | 2.464 | 2 | 389.878 | 432.723 | 2.294 |
| 3 | 442.312 | 387.449 | 2.115 | 3 | 432.723 | 403.710 | 2.130 |
| 4 | 387.449 | 418.996 | 2.316 | 4 | 403.710 | 418.198 | 2.212 |
| 5 | 418.996 | 394.435 | 2.159 | 5 | 418.198 | 409.326 | 2.162 |
| … | … | … | … | … | … | … | … |
| 28 | 403.101 | 403.097 | 2.215 | 18 | 412.399 | 412.406 | 2.179 |
| 29 | 403.097 | 403.100 | 2.215 | 19 | 412.406 | 412.402 | 2.179 |
| 30 | 403.100 | 403.098 | 2.215 | 20 | 412.402 | 412.404 | 2.179 |
| 31 | 403.098 | 403.099 | 2.215 | 21 | 412.404 | 412.403 | 2.179 |
| 32 | 403.099 | 403.099 | 2.215 | 22 | 412.403 | 412.404 | 2.179 |

**Table 1.** Intermediate results in iterative computation of $\lambda_1$ and $n_1(\lambda_1)$.

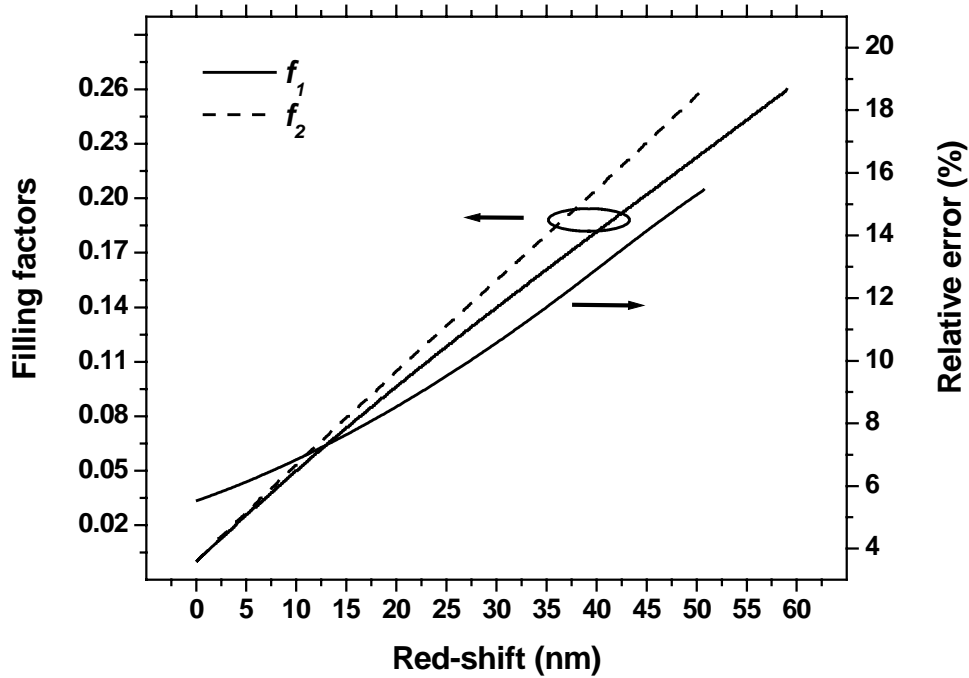

**Fig. 6.** Filling factors and relative error vs. red-shift for ZnO opal infiltrated with GaN.

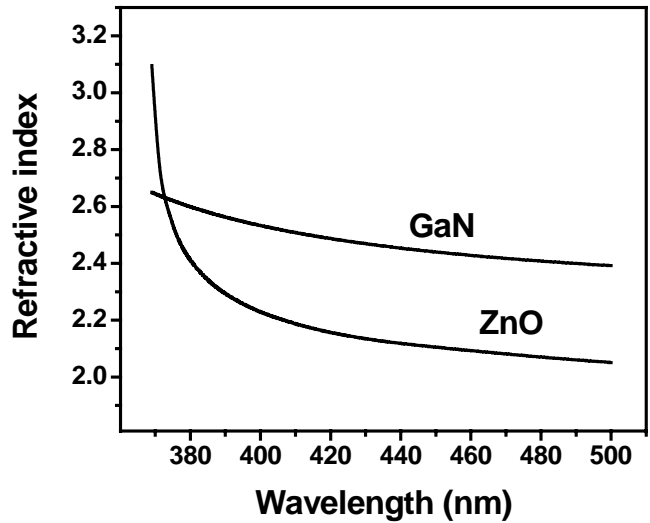

**Fig. 7.** Refractive indices of gallium nitride and zinc oxide.

Figure 7 shows the refractive indices of GaN [24] and ZnO [25, 26]. In the visible spectrum the refractive index of GaN is higher than that of ZnO.

Analyzing (4) and (9) one can see that the relative error mostly depends on the diameter of the balls and the difference between refractive indices of the host and infiltrated materials. The decrease of the refractive index contrast increases the relative error between filling factors. For silica opal infiltrated with ZnO the difference between refractive indices is about 0.5 and more in the visible spectrum, while for ZnO opal infiltrated with GaN it is less than 0.35 (Figure 7). As a result, the relative error for infiltrated ZnO opal is higher. It should be noted, however, that for the visible spectral range, the relative error between filling factors does not exceed 17.5 % even at complete infiltration.

## V. Conclusion

Two equations for the filling factor estimation of infiltrated ZnO in silica opal and GaN in ZnO opal have been considered. The first equation is based on effective medium approximation, while the second one – on Maxwell-Garnett approximation.

Filling factor equations can be simplified for silica opal because of its week dependence of refractive index on wavelength. However, the filling factor equations for opal comprising ZnO balls requires a numerical solution.

The comparison between filling factors shows that both of them can be equally used for quantity estimation of infiltrated material inside opal matrix. However, at complete infiltration the relative error between filling factors for ZnO opal infiltrated with GaN may exceed 15% due to small refractive index contrast between zinc oxide balls and infiltrated gallium nitride nanoparticles.

**Acknowledgements**

This work is supported by the Korea Science and Engineering Foundation through the Quantum-functional Semiconductor Research Center, and by the research program and fund of Dongguk University, 2005.

______________________

Corresponding author:   S. M. Abrarov

Emails:                 absanj@yahoo.co.uk
                        abrarov@dongguk.edu